\documentclass[
twocolumn, amsmath, amssymb, floatfix,
nofootinbib,wrapfloat byrevtex]{revtex4}

\usepackage{amsmath,mathrsfs,graphicx,epstopdf,placeins,float}
\usepackage{booktabs}
\usepackage[unicode]{hyperref}
\usepackage{multirow}
\usepackage{pstricks}
\usepackage{bbm}
\usepackage{epsfig}

\newcommand{\be}{\begin{equation}}
\newcommand{\ee}{\end{equation}}
\newcommand{\ba}{\begin{eqnarray}}
\newcommand{\ea}{\end{eqnarray}}
\newcommand{\p}{\prime}

\begin{document}

\title[ ]{Remarks on Replica Method and Sachdev-Ye-Kitaev Model}

\author{ Ming Chen}
\email{m.chen3@uqconnect.edu.au}
\author{ Yao-Zhong Zhang}
\email{yzz@maths.uq.edu.au}
\affiliation{School of Mathematics and Physics, The University of Queensland,
Brisbane, QLD 4072, Australia}

\date{\today $\vphantom{\bigg|_{\bigg|}^|}$}

\begin{abstract}
Through tracing back to EA/SK models, we disentangle the construction logic of SYK model. During the construction, we recall the technical essence of replica method. Then we point out the redundance of the flavor group and the slave indices as well as the color group via the generalization from 2-index SY model to its 4-index version and the simplification of the latter in terms of Majorana fermions. Spectacularly, such a simplification reserves the system's self-consistency condition. Getting rid of the redundance, the model itself and its effective action have the same $O(N_s)$ structural symmetry, while it realizes $AdS_2/CFT_1$ holography by the emergent conformal symmetry. We also clarify the model transmutation to matrix model and argue that the disorder-averaged SYK model naturally preserves a holographic nature. Based on the transmutation and the correspondence with vector model, SYK model is formally a hybrid of these two models. Its quantum and semiclassical properties are also discussed respectively.

\vspace{1em}

DOI:                      \hfill
\end{abstract}

\maketitle
\section{Introduction: From EA/SK model to SYK model}
Gauge/gravity duality \cite{98maldasyk}-\cite{maldasyk} has been observed within holographic metals \cite{science2}, and this has make spectacular inspirations in holography to 4-index Sachdev-Ye (SY) model \cite{syhol}-\cite{syentropy}, as well as the simplified Sachdev-Ye-Kitaev (SYK) model \cite{sykvedio}-\cite{polr}.\\

The original 2-index SY model is an interaction many-body spin system that can reliably describe the gapless mean-field spin liquids \cite{sy}-\cite{geo}, which can be traced back to the Edwards-Anderson (EA) model \cite{edward} or Sherrington-Kirkpatrick (SK) model \cite{sk}\cite{replicaaj} about the replica theory of spin glasses.\\

Complying with Mermin-Wagner (MW) conclusion \cite{12dnorder}, both SY and EA/SK models are obedient to the condition where the disordered spins are interacting through all-to-all or infinite-ranged random exchange couplings which are independently drawn from Gaussian distribution.\\

To deal with such disordered systems with the quenched random couplings functionally, the replica method has been introduced playing the role of the mean-field theory \cite{edward}\cite{pastur}\cite{tala}.\\

The original motivation of such a method is to integrating out the spins by altering the integral order between spins and couplings. Phrased another way, replica method is a technique to complete the mixed functional integrals of spins and couplings in the way of isolating them. Moreover, the replica index ($m$) is valid in all its values and can be taken into infinitesimal to get the approximated solutions \cite{edward}\cite{aref}.\\

Thus technically, the replica method can be achieved through the following identity in the sense of averaging the free energy \cite{sk},
\begin{eqnarray}\label{syk21}
\textrm{ln} x=\lim_{_{m\rightarrow 0}} \frac{x^{m}-1}{m}.
\end{eqnarray}

However, for the replica-diagonal situation \cite{sy}, the replica method can be more tricky \cite{kitaev},
\begin{eqnarray}\label{syk22}
\textrm{ln} x=\lim_{_{m\rightarrow 0}} \frac{\textrm{ln} x^{m}}{m},
\end{eqnarray}
which makes the replica be kind of trivial \cite{syksigma}.\\

Meanwhile, in the sense of mean-field theory, the replica is formally equivalent with slave fermions \cite{baska}\cite{wenm}\renewcommand{\thefootnote}{\fnsymbol{footnote}}\footnote[3]{Chapter 9.1}\renewcommand{\thefootnote}{\arabic{footnote}}\cite{syads} for such disordered systems. This allows one to transform 2-index SY model to 4-index form \cite{sy}\cite{syhol}.\\

In detail, generalized from EA/SK model, the 2-index SY model is governed by the following action,
\begin{eqnarray}\label{syk23}
S =S_{_B}-\int d\tau H_{\textrm{int}},
\end{eqnarray}
where $S_{_B}$ denotes the Berry phase (BP) term, coming from coherent-state path integral (CSPI) \cite{coherents}\cite{wenm}\renewcommand{\thefootnote}{\fnsymbol{footnote}}\footnote[2]{Chapter 2.3}\renewcommand{\thefootnote}{\arabic{footnote}}, to characterize the single-spin kinematic property, and the interaction Hamiltonian of the system is,
\begin{eqnarray}\label{syk24}
H_{\textrm{int}}\sim \frac{1}{\sqrt{N_s N_f}}\sum^{N_s}_{i,j}\sum^{N_f}_{m,n} \mathcal{J}_{ij}\hat{S}^{m}_{i}(\tau)\hat{S}^{n}_{j}(\tau^{\p}),
\end{eqnarray}
with $\hat{S}^{m}$ the generators in the representation of flavor group $SU(N_f)$, $N_s$ the number of sites, and the sum over sites $(i,j)$ extends over all $N_s \rightarrow \infty$ sites.\\

Clearly, besides the invariance under $SU(N_f)$ symmetry, the Hamiltonian is also invariant in the spin-rotation symmetry which can be expressed as $\mathbb{Z}_2^{N_s}$ \cite{tala}\cite{tala2}. Because $\mathbb{Z}_2$ is the center of $O(N)$, $\mathbb{Z}_2^{N_s}$ can be generalized to $O(N_s)$ symmetry.\\

The quenched couplings $\mathcal{J}_{ij}$ are all randomly taken from a Gaussian probability distribution which are all infinite-range and independent,
\begin{eqnarray}\label{syk25}
\mathcal{P}(\mathcal{J}_{ij})=\frac{1}{\sigma\sqrt{2\pi}}exp \big( -\frac{1}{2}\frac{\mathcal{J}_{ij}^2}{\sigma^2} \big),
\end{eqnarray}
with zero mean $\overline{\mathcal{J}_{ij}}=0$ and variance $\sigma^2=\overline{\mathcal{J}_{ij}^2}=\frac{J^2}{N_s}$ where $J$ is the characteristic energy scale of the system.\\

They play their role during the average on the spins' disorders,
\begin{eqnarray}\label{syk26}
\langle \cdots \rangle _{\mathcal{J}}\sim \int^{+\infty}_{-\infty} \prod^N_{ij} \big[ d\mathcal{J}_{ij} \mathcal{P}(\mathcal{J}_{ij})\big]\cdots .
\end{eqnarray}

Together with Eq.\eqref{syk24}, technically, if we take the flavor indices $(m,n)$ as the replica indices and make one Gaussian integration (or namely the Hubbard-Stratonovich Transformations \cite{hirsch}-\cite{hagen}) to integrate out the couplings, then we can arrive at the replica-form 4-index effective action. Analogous to Eq.\eqref{syk23}, with the Hamiltonian in the following,
\begin{eqnarray}\label{syk27}
H_{\textrm{int}}\sim \frac{J^2 3!}{2N^3_s} \sum^{N_f}_{m,n} \int^{\beta}_0 d \tau d \tau^{\p}  \hat{S}^m(\tau) \hat{S}^n(\tau^{\p})\hat{S}^m(\tau) \hat{S}^n(\tau^{\p}).
\end{eqnarray}

Clearly, such kind of mean-field limit reduce the disordered spin systems into the single-site problem with $m$($n$) replicas. This equals with the slave-fermion expression in terms of spinon operators $f_i^m$,
\begin{eqnarray}\label{syk2add1}
\hat{S}^m_i\sim \sum^{N_f}_{m} f^{\dag m}_{ i}f^{m}_i.
\end{eqnarray}

That is,
\begin{eqnarray}\label{syk28}
H_{\textrm{int}}\sim \frac{J^2 3!}{2N^3_s} \sum^{N_s}_{i}\sum^{N_f}_{m} \int_0^{\beta} d\tau d\tau^{\p} |f^{\dag m}_{i} f^m_{i}|^4.
\end{eqnarray}

With spinon operators, the kinematic property can be represented differently from $S_{_B}$ \cite{syads},
\ba\label{syk29}
S_{_B}  \sim \sum^{N_s}_{i}\sum^{N_f}_{m} \int_0^{\beta}  d\tau (- f^{\dag m}_{i} \frac{d}{d\tau}f^m_{i}).
\ea

Combination of Eq.\eqref{syk28} and Eq.\eqref{syk29} gives the full replica-form effective action,
\begin{flalign}\label{syk210}
& S_{\textrm{eff}}=\sum^{N_s}_{i}\sum^{N_f}_{m} \bigg\{ \int_0^{\beta}  d\tau (- f^{\dag m}_{i} \frac{d}{d\tau}f^m_{i}) - & \nonumber \\
& \;\;\;\;\;\;\;\;\;\;\;\;\;\;\;\;\;\;\;\;\;\;\;\;\;\;\;\;\;\;\;\;\;\;\;\;\;\;   \frac{J^2 3!}{2N^3_s} \int_0^{\beta} d\tau d\tau^{\p} |f^{\dag m}_{i} f^m_{i}|^4 \bigg\}. &
\end{flalign}

The saddle points of the effective action, e.g., the Green functions of the disorder-averaged spin liquids system, are all replica-diagonal, thus the replica indices can be dropped. This is consistent with the fact that the replica is trivial in spin liquids system \cite{sy}\cite{kitaev}\cite{syksigma}, we can ignore it at the beginning and simplify and reduce the effective action back to the 4-index form before the disorder average,
\begin{flalign}\label{syk211}
& S \sim \sum^{N_s}_{j} \int_0^{\beta}  d\tau (-f^{\dag}_{j} \frac{d}{d\tau}f_{j}) - & \nonumber \\
& \;\;\;\;\;\;\;\;\;\;\;\;\;\;\;\;\;\;\; \sum^{N_s}_{j,k,l,m} \int_0^{\beta} d\tau d\tau^{\p} \mathcal{J}_{jk;lm}(f^{\dag}_{j} f_{k})(f^{\dag}_{l} f_{m}). &
\end{flalign}

On the one hand, under such simplification and reduction, the replica indices and the flavor group of spins \cite{sy} as well as the color group in CSPI \cite{coherents} are actually all redundant for the model itself.\\

On the other hand, with such a fermionic spinon representation, the system will describe the spin liquids since fermions are naturally incompatible at one single site and the spin glasses are manifestly excluded. Compared with the ordered system which can be described by a semiclassical procedure, the spin liquids system are intrinsically quantum mechanical \cite{sy}. \\

For such a quantum system, its canonical quantization condition is expressed by fermions' anti-commuting relations,
\ba\label{syk214}
\{f^{\dag}_{j}, f_{k} \}=\delta_{jk},\;\; \{f^{\dag}_{j}, f^{\dag}_{k} \}=0,\;\; \{f_{j}, f_{k} \}=0.
\ea

If we transform Eq.\eqref{syk211} based on Eq.\eqref{syk214}, there will be an additional Kronecker $\delta$-function term in the Hamiltonian,
\ba\label{syk215}
H_{\textrm{int}} \sim \sum^{N_s}_{j,k,l,m} \mathcal{J}_{jk;lm} f^{\dag}_{j} f^{\dag}_{k} f_{l} f_{m} -\delta_{kl}\sum^{N_s}_{k,l} \mathcal{J}(f^{\dag}_{k} f_{l}).
\ea

Such an addition protects the Hamiltonian's globally $U(1)$ symmetry and the corresponding globally conserved charge, which are manifestly shown from the change of the spinon number constraint,
\ba\label{syk212}
\sum^{N_f}_{m} f^{\dag m}_{ i}f^{m}_i=N_f \mathcal{Q} \Leftrightarrow \sum^{N_s}_{i} \mathcal{J}(f^{\dag}_{ i}f_i)=N_s \mathcal{Q},
\ea
with $0<\mathcal{Q}<1$ the $U(1)$ charge density which has the meaning of order parameter ($\mathcal{Q}^2$) of spin glasses in EA model \cite{edward}.\\

This $U(1)$ symmetry plays a vital role in the relation between SY model and universal low temperature theory of charged black holes from the Bekenstein-Hawking Entropy prospective \cite{syentropy}\cite{sykcomplex}\cite{syhol}.\\

However, the addition is actually the chemical potential within the spinon representation and this will destroy the self-consistency condition of the spin liquids system. The reason is that the quenched order in the spin-represented SY model is generated by the gapless ``heat bath'' which is self-consistently generated internally by the other spins, the model can be taken as a canonical ensemble. With this potential, the canonical ensemble will transformed to grand canonical ensemble.\\

In order to reserve the self-consistency condition and keep the model simple, one way is replacing spinons in Eq.\eqref{syk211} by Majorana fermions (MFs) since they are exactly the same as their Hermitian correspondence,
\ba\label{syk216}
\chi^{\dag}=\chi, \;\;\;\;\; \{ \chi_{_j}, \chi_{_k} \}=\delta_{jk}.
\ea

In this way, we are arriving at SYK model,
\begin{flalign}\label{syk217}
& S \sim \sum^{N_s}_{j} \int_0^{\beta}  d\tau (-\chi_{_j} \frac{d}{d\tau}\chi_{_j}) - & \nonumber \\
& \;\;\;\;\;\;\;\;\;\;\;\;\;\;\;\;\;\;\; \sum^{N_s}_{j,k,l,m} \int_0^{\beta} d\tau d\tau^{\p} \mathcal{J}_{jklm}(\chi_{_j} \chi_{_k}\chi_{_l} \chi_{_m}). &
\end{flalign}

In such Majorana representation, the heat bath quenches the order comes from the self-consistently effective MF field \cite{sykvedio2}\cite{sykvedio}\cite{kitaev},
\ba\label{syk218}
\mathcal{O}_j=-i\frac{\partial H_{\textrm{int}}(\chi)}{\partial \chi_{_j}} \sim \sum^{N_s}_{k,l,m} \mathcal{J}_{klm}\chi_{_k} \chi_{_l} \chi_{_m}.
\ea

Besides the ignored $U(1)$ symmetry, SYK model also differs in symmetry from the original SY model. Without replica, the $SU(N_f)$ symmetry is directly cancelled while the $O(N_s)$ symmetry is reserved by particle-hole symmetry based on Eq.\eqref{syk216}.\\

By now, for the model itself, SYK model is a self-consistently quenched disordered quantum mechanical spin liquids system of $N_s$ all-to-all interaction MFs invariant under the $O(N_s)$ symmetry.\\

For its algebra, we can normalize the algebra basis and rescale the anti-commuting relation in Eq.\eqref{syk216} to $\{ \chi_{_j}, \chi_{_k} \}\!\!=2\delta_{jk}$ with $(\chi_{_j})^2=1$. Therefore, the algebra for MF operators must be Clifford algebra with orthogonal basis,
\ba\label{ford}
\big\{ \{ \chi_{{_j}_1}\chi_{{_j}_2}\cdots \chi_{{_j}_q} \}  \big| 1\!\leq \!j_1\! <\! j_2\! <\!\! \cdots \!\!< \!j_q \!\leq \!N_s, 0 \!\leq\! q \!\leq \!N_s \big \}
\ea
and the total dimension of Clifford Algebra is $[\mathcal{CL}]=2^{N_s}$, while the dimension of Hilbert Space is $[\mathcal{HIL}]=2^{N_s/2}$ \cite{west}\cite{mik}\cite{jose}.\\

The different dimensions between Clifford and Hilbert spaces explain the constraint of Eq.\eqref{syk212}. That is, when we change from Hilbert space (2-index: one spin state per site or two fermions per site) to Clifford space (4-index: one fermion per site), we have to impose the fermion number constraint to make sure the equivalence between the varied Hamiltonians.\\

To be noted, without the globally $U(1)$ symmetry, SYK model finds another way to dual with black hole, i.e., the ``shock wave'' which will be showed up in the later Out-of-Time-Order Correlators (OTOCs).

\section{Effective action and Large-$N_s$ structure}
Similar with the procedure in 4-index SY model (Eq.\eqref{syk25}-Eq.\eqref{syk210}), the replica-form and disordered partition function of SYK model is \cite{kitaev},
\begin{flalign}\label{syk219}
& \langle Z^m(\mathcal{J}) \rangle =\int^{+\infty}_{-\infty} \mathcal{D}\mathcal{B} Z^m(\sigma \mathcal{B})= & \nonumber \\
&  \int^{+\infty}_{-\infty} \!\!\!\! \prod^{N_s}_{j,k,l,m} \big[\frac{1}{\sqrt{2\pi}}\textrm{exp}(-\frac{1}{2}\mathcal{B}_{jklm}^2)d\mathcal{B}_{jklm} \big] \times & \nonumber \\
& \int^{+\infty}_{-\infty}\mathcal{D}\chi  \bigg[ \;\textrm{Tr} \; \textrm{T} \big\{ \textrm{exp}\big( \sum^m \int^{\beta}_0 \!\!\!\! d\tau ( -\frac{1}{2} \sum^{N_s}_{j=1} \chi^{m}_{_j} \frac{d}{d\tau}\chi^{m}_{_j}) \big)- & \nonumber \\
&  \textrm{exp} \big( \sum^m \int^{\beta}_0 \!\!\!\! d\tau d\tau^{\p} (\!\! \sum^{N_s}_{j,k,l,m} \!\!\!\! \sigma \mathcal{B}_{jklm} \chi^{m}_{_j} \chi^{m}_{_k} \chi^{m}_{_l} \chi^{m}_{_m} )  \big) \!\! \big\} \bigg]. &
\end{flalign}
with $\mathcal{B}_{jklm}=\frac{\mathcal{J}_{jklm}}{\sigma}$. In this way, the order between the integration on MFs and couplings can be altered, we can thus first integrate out the couplings,
\begin{flalign}\label{syk220}
& \langle Z^m \rangle=\!\!\!\! \int^{+\infty}_{-\infty} \!\!\!\!\!\!\!\! \mathcal{D}\chi \textrm{Tr} \; \textrm{T} \bigg\{ \textrm{exp} \big( -\frac{1}{2} \sum^m \sum^{N_s}_{j=1} \int^{\beta}_0 \!\!\!\! d\tau \chi^{m}_{_j} \frac{d}{d\tau}\chi^{m}_{_j} +  & \nonumber \\
& \sigma^2 \!\!\!\! \sum^{N_s}_{j,k,l,m} (\sum^m \int^{\beta}_0 d\tau d\tau^{\p} (\chi^{m}_{_j}(\tau) \chi^{m}_{_k}(\tau^{\p}) \chi^{m}_{_l}(\tau) \chi^{m}_{_m}(\tau^{\p}) )^2 \big)  \bigg\}. &
\end{flalign}

It is interesting to see the disorder-averaged replica-form Hamiltonian of this model,
\ba\label{syk221}
\mathcal{H}\sim \sigma^2  \sum^{N_s}_{j,k,l,m}  \sum^m \big( \chi^{m}_{_j}(\tau) \chi^{m}_{_k}(\tau) \chi^{m}_{_l}(\tau) \chi^{m}_{_m}(\tau) \big)^2
\ea

If we take the slave fermions in Eq.\eqref{syk2add1} as color-singlet quark bilinears ($M_{jk} \!\! \sim\!\! \sum^{N_c}_{_{jk}} \!\! \bar{q}_{_j}  q_{_k}$) \cite{wein}-\cite{ming}, such a form of Hamiltonian actually resembles matrix model formally \cite{thooft}-\cite{coleman}\renewcommand{\thefootnote}{\fnsymbol{footnote}}\footnote[2]{Chapter 8}\renewcommand{\thefootnote}{\arabic{footnote}}.\\

This can be seen from bilinears' Green functions based on the above Hamiltonian, like that, the 4-point vertex,
\begin{flalign}\label{syk222}
& \langle \textrm{Tr} M^4\rangle  \sim \langle \textrm{Tr} \textrm{\textrm{T}} \big\{ \sum^{N_s}_{j,k;l,m} \big( [\chi^{m}_{_j}(\tau) \chi^{m}_{_k}(\tau)] [\chi^{m}_{_l}(\tau) \chi^{m}_{_m}(\tau)] \big)^2\big\} \rangle  & \nonumber \\
& \;\;\;\;\;\;\;\;\;\;\;\; \sim (2N_s^3+N_s). &
\end{flalign}

From the one side, this reveals the advantages of obvious formal simplicity in the calculation of Green functions by replica method compared with diagrammatical operations.\\

From the other side, when we deal with large-$N_s$ limit, the first contribution is dominant undoubtedly. This provides the guide to determine the variance. To make the 4-point Green function exact, the variance can be tuned as following,
\ba\label{syk223}
\sigma^2=\overline{\mathcal{J}^2_{jklm}}=\frac{J^2 3!}{N^{3}}.
\ea

This makes the model much simpler than the tensor model \cite{gurau}-\cite{syktrick}.\\

Together with the zero mean which is naturally in the random-site situation \cite{edward}, the mean and variance of certain probability distribution will be,
\begin{eqnarray}\label{syk224}
\left\{
\begin{array}{ll} 
\ \overline{\mathcal{J}_{jklm}}=0,\\ \\
\ \overline{\mathcal{J}^2_{jklm}}=\frac{J^23!}{N^3}.
\end{array}
\right.
\end{eqnarray}

They reduce the diagrams or the contributions of the model in the large-$N_s$ limit to a subset which can be summed exactly, thus guarantees the theory to be solvable
in principle.\\

It is clear that the exact form of probability distribution is not essential during the previous calculation. However, when we trace back to the EA/SK model, the disordered system is naturally approximated by Gaussian distribution in the large-$N_s$ limit, and the exact Gaussian distribution of the SYK model can be expressed in the following,
\ba\label{syk225}
\mathcal{P}(\mathcal{J}_{ijkl})= \sqrt{\frac{N^3}{12\pi J^2}}\textrm{exp} \big( -\frac{N^3 \mathcal{J}^2_{jklm}}{12J^2} \big).
\ea

We then can continued from Eq.\eqref{syk220} to integrate out the MFs and arrive at the exact replica-free effective action,
\begin{eqnarray}\label{syk226}
\left\{
\begin{array}{lll} 
\ \langle Z\rangle=\int^{+\infty}_{-\infty} \mathcal{D}G \mathcal{D}\Sigma \;\; \textrm{exp}(S_{\textrm{eff}});\\ \\
\ S_{\textrm{eff}} [G, \Sigma] = N_s \mathcal{S} =N_s \bigg\{ \textrm{ln} \;\textrm{Pf}\big(-\partial_{\tau}-\Sigma(\tau,\tau^{\p})\big)+ \\ \\
\ \frac{1}{2} \int_{\tau>\tau^{\p}}\int^{\beta}_0 d\tau d\tau^{\p} \bigg( \frac{J^2}{4} G^4(\tau,\tau^{\p})-  \Sigma(\tau,\tau^{\p})G(\tau,\tau^{\p}) \bigg)  \bigg\}.
\end{array}
\right.
\end{eqnarray}

In the large-$N_s$ limit, the Dyson equations or Born series \cite{paul} can be calculated through the saddle points of the effective action \cite{sy}\cite{kitaev},
\begin{eqnarray}\label{syk227}
\left\{
\begin{array}{ll} 
\ G(\tau,\tau^{\p})=\frac{1}{-\partial_{\tau}-\Sigma(\tau,\tau^{\p})} \rightarrow \frac{1}{-i\omega-\Sigma(i\omega)};\\ \\
\ \Sigma(\tau,\tau^{\p})=J^2 G^3(\tau,\tau^{\p}),
\end{array}
\right.
\end{eqnarray}

Clearly, the saddle point values of $G$ and $\Sigma$ are respectively the 2-point Green functions ($G \!\! \sim\!\! \sum^{N_s}_i \!\! \chi_{_i} (\tau) \chi_{_i}(\tau^{\p})$) and the self-energy in such dynamic mean-field approximation.\\

In fact, we can eliminate the self-energy and have just the Green functions which are functions of two time variables in the effective action. Similar with the color-singlet quark bilinears which are invariant under $O(N_c)$ symmetry, the Green functions are invariant under the $O(N_s)$ symmetry. Therefore, the effective is also $O(N_s)$ invariant.\\

That is, for both the model itself and its effective action, SYK model preserves the $O(N_s)$ symmetry.

\section{Holography Approach and Model Transmutation}
In the IR scaling limit, the high frequency term can be separated \cite{syktrick}\cite{sykcomplex}\cite{syentropy}\cite{kitaev}, and the saddle points display an emergent time-reparameterizations symmetry,
\begin{eqnarray}\label{syk228}
\left\{
\begin{array}{ll} 
\ G(\tau,\tau^{\p})=\big[ \frac{\partial f(\tau)}{\partial \tau} \frac{\partial f(\tau^{\p})}{\partial \tau^{\p}} \big]^{1/4} G(\sigma,\sigma^{\p});\\ \\
\ \Sigma(\tau,\tau^{\p})=\big[ \frac{\partial f(\tau)}{\partial \tau} \frac{\partial f(\tau^{\p})}{\partial \tau^{\p}} \big]^{3/4} \Sigma(\sigma,\sigma^{\p}),
\end{array}
\right.
\end{eqnarray}
where $\sigma=f(\tau)\sim e^{i \phi (\tau)}$. This is the critical conformal property of SYK's solution because it obviously preserves diffeomorphisms and Weyl invariance in Euclidean space (it is common to switch to Minkowski space by Wick rotation, vice versa).\\

We can then choose the $SL(2, \mathbb{R})$ symmetry for such IR regime to fit the conformal transformations, i.e.,
\ba\label{syk229}
f(\tau)=\frac{a\tau+b}{c\tau+d}, \; a,b,c,d\in \mathbb{R}, \; ad-bc=1.
\ea

Clearly, such a symmetry will be spontaneously broken by the separated UV effects.\\

In the meantime, the characteristic energy scale $J$ of the system has the dimension $[J]=1$. As a result, the model will be strongly coupled in the IR regime, and such a strong-coupling property is enhanced by an effective coupling constant $\beta J$ with $\beta \sim \frac{1}{T}$ in the low-temperature limit.\\

Therefore, the effective theory of SYK model in the IR regime is an emergent strongly coupled $CFT_1$ theory and this paves the way to holography.\\

First of all, compared with weakly-coupled vector $O(N)$ models \cite{sim}, the effective strong-coupling property is essential to trust the gravity approximation from the side of gauge theory \cite{maldasyk}\cite{cmt}.\\

Secondly, according to holography theory, the emergent $CFT_1$ corresponds to $AdS_2$ space because the $AdS_{_D}$ space is bounded by Minkowski space $\mathbb{R}^{1,D-2}$,
\ba\label{syk231}
\partial AdS_{_D}=\mathbb{R}^{1,D-2}.
\ea

This is also confirmed by the fact that both effective theory of SYK model in the IR regime and $AdS_2$ space preserve the $SL(2,\mathbb{R})$ gauge-like symmetry \cite{sl2rak}.\\

As above, $AdS_2$ space has a holographic correspondence with SYK-$CFT_1$ field theory.\\

Furthermore, $AdS_2$ space can be characterized by Schwarzian action which is the simplest action of $SL(2,\mathbb{R})$ symmetry and related to black holes.\\

This implies that, SYK-$CFT_1$ field theory can also be described by the Schwarzian action, and the field $f(\tau)$ is referred to as the soft mode of bulk gravity in $AdS_2$ space \cite{jack}-\cite{ads5}. As a result, SYK-$CFT_1$ indeed gives the gravitational mode \cite{tay} and relates to black holes.\\

For black holes, the semiclassical theory of gravity gives some amazing results \cite{proch}-\cite{hawk4}. Therefore, if SYK-$CFT_1$ is about ``quantum gravity'', there muse be some universal behaviors of black holes from it.\\

Greatly, black holes have one kind of gravitational mode namely the ``shock wave'' \cite{dray}\cite{horo} which show ups in the aforementioned OTOCs \cite{larkin}-\cite{ads2}.\\

Instead of requiring a complete theory of quantum gravity, OTOCs allow the comparison between black holes and the conventional many-body systems. The black hole OTOCs at early times (but after all 2-point correlators have decayed) have some characteristic properties that reflect the physics near the horizon \cite{sykvedio3}.\\

One significant feature of OTOCs is their time dependence,
\ba\label{syk232}
\frac{G(0,\tau;0,\tau)}{G(0,0)G(\tau,\tau)} \propto e^{\alpha \tau},
\ea
where $\alpha=\frac{2\pi}{\beta}$ denotes the Lyapunov exponent and $\beta$ here is the reciprocal of Hawking temperature $T$.\\

Usually, the Lyapunov exponent is smaller. In fact, there is general relation $\alpha \leq 2\pi / \beta$ for any quantum system at temperature $T$ \cite{ads3}. However, it is rather difficult to satisfy, e.g., the 2-index SY model.\\

For SYK-$CFT_1$ \cite{polr},
\ba\label{syk233}
\frac{G(0,\tau;0,\tau)}{G(0,0)G(\tau,\tau)} \propto \frac{\beta J}{N_s}e^{\alpha \tau}.
\ea

That is,
\ba\label{otoclya}
\alpha \approx \frac{2\pi}{\beta}, \; \textrm{if}\; \beta J\gg 1.
\ea

As a result, the SYK-$CFT_1$ has the maximum Lyapunov exponent thus matches the black hole result comparatively well.\\

Such a match again verifies the connection between the effective SYK model and 2-dimensional bulk gravity \cite{commsyk}\cite{kitaev}\cite{syktrick}\cite{sykcomplex}, and identifies the soft mode in SYK model with 't Hooft's gravitational modes \cite{ads5}\cite{ads6}-\cite{ads8}.\\

This is the also reason why the holography must be achieved in the 4-index form. Actually, the reason why we have to turn to such 4-index form can also be seen from Eq.\eqref{syk221} and Eq.\eqref{syk222}.\\

As we have argued, the disorder-averaged Hamiltonian resembles the matrix model formally through the structural similarity between Green functions and quark bilinears.\\

Furthermore, since the Green functions are of two variables, the effective action of Green functions can be taken as a field theory defined on a 2-dimensional space. This again supports the structural similarity between the disorder-averaged action and the 2-dimensional large-$N_c$ matrix model. What's more, when both of them are strongly coupled, they are all gapless and having no quasiparticles.\\

In other words, based on the disorder average, Eq.\eqref{syk221} shows certain kind of model transmutation between these two models. Eq.\eqref{syk222}-like Green functions justify such a transmutation because both of the models shares the same mechanism to choose what subsets of the theories contribute effectively to the Green functions.\\

Like that, for the matrix model, the foundation of holography lies in the fact that its Green functions in the strongly coupled situation can be expressed in terms of the topology of Riemann surfaces in which the chosen subset (planar diagrams) can be embedded \cite{topoexp1}-\cite{topoexp3},
\ba\label{largesyk}
G \sim \sum^{\infty}_{h,b=0} \big(\frac{1}{N_c} \big)^{-\chi} \sum^{\infty}_{n=0} c_{h,b;n} \lambda^n,
\ea
where $\chi=2-2h-b$ is the topologically invariant Euler characteristic number which characterizes the topology of the manifolds.\\

This foundation builds the relation with string theory \cite{tong}\cite{sykwstr}, especially the Type II$B$ superstring theory which has an $SL(2,\mathbb{R})$ symmetry that when quantized reduces to $SL(2,\mathbb{Z})$. Meanwhile, in addition to the global symmetry group $SU(2;2|4)$, the Super Yang-Mills (SYM) theory also has the (conjectured) $S$-duality symmetry group $SL(2,\mathbb{Z})$.\\

Very overall speaking, after checking thoroughly through the field maps and Green functions between the model and string theory, the symmetry accordance generalizes to general holographic correspondence between the quantum gravity on an $AdS$ background and conformally-invariant non-gravitational field theory defined on such $AdS$ boundary \cite{98maldasyk}-\cite{maldasyk}, e.g.,
\begin{flalign}\label{ads2bn4}
& (\textrm{Type}\; \textrm{II}B\textrm{ superstring theory on} \; AdS_5\times S^5)  \equiv & \nonumber \\
& (\mathcal{N}=4\; \textrm{SYM  in} \;4\!-\!\textrm{dimensional spacetime}). &
\end{flalign}

Compared with the holography approach of SYK model, these two models share the similar approach towards holography: the same symmetry between two theories implies certain correspondence.\\

In this meaning, although these two models varies a lot in details, such a transmutation indeed naturally allows the disorder-averaged form of SYK model to share the holography property of the matrix model.\\

At the same time, the kinematic BP term is a general effect within CSPI. It is a geometric phase which represents the amount of frustration of assigning or parallel transporting a common phase to all the coherent states.\\

Moreover, the concepts of parallel transportation and the frustration in parallel transportation are very important because both electromagnetic field and gravitational field are generalized BPs \cite{wenm}\renewcommand{\thefootnote}{\fnsymbol{footnote}}\footnote[2]{Chapter 2.3.3}\renewcommand{\thefootnote}{\arabic{footnote}}.\\

Like that, SY's BP is kind of magnetic flux \cite{coherents}. Thus we should ignore it towards holography and, fortunately, this is reasonable at low energies.\\

In a word, the disorder-averaged SYK model as a whole preserves a holographic nature with strong couplings in the IR regime.

\section{Correspondence with Vector Models}
Different from the hidden model transmutation, the effective action which emerges the SYK-$CFT_1$ theory apparently preserves the $O(N_s)$ symmetry. What's more, the large $N_s$ factor in the effective action makes the theory in terms of $G$ and $\Sigma$ be semiclassical.\\

For such semiclassical situation, it is convenient to analyse it from the conventional vector $O(N)$ models.\\

There have been work showing the structural similarity between the effective action and certain kind of nonlinear $\sigma$ model by rewriting Schwarzian in terms of certain vector field \cite{syksigma}. Or more directly, representing the disordered-averaged action through bi-local fields (Green functions) to map the form of vector model \cite{syksusy2}-\cite{antal2}.\\

Here, we will simply go from the linear $\sigma$ model to discuss the semiclassical property of SYK's effective action because the nonlinear $\sigma$ model is just the geometrically generalized linear situation \cite{coleman}\renewcommand{\thefootnote}{\fnsymbol{footnote}}\footnote[2]{Chapter 8}\renewcommand{\thefootnote}{\arabic{footnote}}.\\

The linear $\sigma$-model is a theory of $N$ scalar fields packaged as an $N$ vector field ($\{ \phi^i: \; i=1,\cdots, N; \; i \in \mathbb{Z}\}$) coupled by a $\phi^4$-type interaction \cite{peskin}\renewcommand{\thefootnote}{\fnsymbol{footnote}}\footnote[1]{Chapter 4-Chapter 13}\renewcommand{\thefootnote}{\arabic{footnote}}\cite{paul},
\begin{eqnarray}\label{mf2}
H\!\! \sim \!\! \int \!\! d^{^{D-1}}x \bigg\{ \!\! \sum_i^N  \partial_{\mu}\phi^i \partial^{\mu} \phi^i \!\!-\!\!  \mu^2 \sum_i^N \phi^i\phi^i \!\!+\!\! \lambda \big( \sum_i^N \!\! \phi^i\phi^i \big)^2 \bigg\}
\end{eqnarray}

To make the $\frac{1}{N}$ dependence transparent, we can introduce an auxiliary field $\sigma$,
\begin{eqnarray}\label{rescs}
\left\{
\begin{array}{ll} 
\ S[\phi^i] \!\! \sim \!\! \int \!\! d^{^D}x \bigg\{ \partial_{\mu}\phi^i\partial^{\mu} \phi^i + \frac{N}{g}\sigma^2 - \sigma ( \phi^i\phi^i ) \bigg\},\\ \\
\ \sigma(x^{^{D-1}}) \sim \frac{g}{N}(\phi^i\phi^i \big).
\end{array}
\right.
\end{eqnarray}

Then the effective action is proportional to $N$,
\begin{eqnarray}\label{effsngauss}
S_{_{\textrm{eff}}}(\sigma,N)\sim N \big\{ \textrm{Tr} \; \textrm{ln}(\partial^2+ \sigma)^{-\frac{1}{2}}+\int dx^{^{D-1}} \frac{1}{g}\sigma^2 \big\}.
\end{eqnarray}

Functionally, such an $N$ property can be sought from the replica method based on the definition of $\sigma$, we can treat the index ``i'' of $\phi^i$ fields as the replica index,
\begin{eqnarray}\label{replin}
Z_{_N}[J]\!=\!(Z[J])^{^N}\!\!\!=e^{N\textrm{ln}Z[J]}\sim 1+N\textrm{ln}Z[J]+\mathcal{O}(N^2).
\end{eqnarray}

Because the disconnected diagrams is proportional to higher powers of $N$, $\textrm{ln}Z[J]$ denotes the assemble encoding the fully connected contributions,
\begin{eqnarray}\label{effsnrep}
N\; \textrm{ln}Z[J] \sim N \; S_{_{\textrm{eff}}}(\sigma,g).
\end{eqnarray}

From the replica procedure we can see that, although the auxiliary $\sigma$ field cannot preserve the $\mathbb{Z}_2$ symmetry in the effective action because of spontaneously breaking, it does reserve the reduced $O(N-1)$ structural symmetry itself.\\

However, the $\sigma$ field can be diagonalized by an orthogonal transformation,
\begin{eqnarray}\label{diasig}
\left\{
\begin{array}{ll} 
\ \sigma \delta_{ij}=O^{T}_i \sigma_{ij} O_j,\\ \\
\ O^{T}O=\textbf{1}.
\end{array}
\right.
\end{eqnarray}

This allows us to reduce the replica-diagonal $\sigma$ model to the $\phi^4$-theory with simply $\mathbb{Z}_2$ symmetry, similar with the situation when we deal with SYK model.\\

Spectacularly, for the $\phi^4$-theory, if we shift the scalar field strength,
\begin{eqnarray}\label{sshift}
\phi \mapsto \sqrt{\hbar} \phi,
\end{eqnarray}
then semiclassically,
\begin{eqnarray}\label{shshift}
S[\phi]/\hbar \sim \int d^{^D}\!\! x \; \big[ \partial_{\mu}\phi \partial^{\mu} \phi \!-\! \mu^2 \phi^2 +(\hbar\lambda) \phi^4 \big].
\end{eqnarray}

Clearly, $\hbar$ takes the same position as $\lambda$ in the expansion and this tells us that the effective coupling of the rescaled theory is just $\hbar\lambda$.\\

From the duality between quantum field theory and statistical mechanics, $\hbar$ is dual to the temperature $T$ or $\frac{1}{\beta}$. If we consider the coupling coefficient to be positive dimensional, then the low energy (IR regime) or low temperature will correspond to a strong coupling $J$.\\

Therefore, if we consider certain IR physics with a positive dimensional coupling, the weak combination $\hbar\lambda$ here will correspond to $\frac{1}{\beta J}$ \cite{replicaaj}\cite{kitaev} which is just the reciprocal of the strongly coupling in the effective theory of SYK model.\\

Because the semiclassical expansion in powers of $\hbar$ is generally equivalent with the loop expansion of the field theory \cite{aneesh}, so these discussions make sense in any field theory.\\

Of course, it is farfetched to make the exact correspondence without exact maps of fields and Green functions and so on. However, just like the couping correspondence between matrix model and string theory, this correspondence at least displays some semiclassical property of the effective action of SYK model \cite{sykhoax}.

\section{Summary and Outlook}
To summarise, during the brief review from EA/SK models to SYK model, we firstly recall the essence of replica method. That is, it is a technique to complete the mixed functional integrals by isolating them with the help of replica indices. For the replica-diagonal SYK model, such an essence is obvious and even trivial.\\

Secondly, from the generalization and simplification from 2-index SY model to 4-index version, technically speaking, the flavor group and the slave fermions play the equivalent role in replica method formally. Together with the color group in CSPI, these three group indices are all redundant if there is replica procedure. Thus, the flavor group in the 2-index SY model can be ignored and simplified to 4-index SYK model.\\

Particularly, the simplification from 4-index SY model to SYK model reserves the system's self-consistency condition through the anti-commuting relations of MFs.\\

Compared with the diagrammatical approach in the calculation of Green functions, replica method takes the advantage of formal simplicity in the determination of dominant contributions to the model, and such an advantage is enhanced through the variance of certain probability distribution after disorder-averaging.\\

Disorder averaging procedure is one formal progress to achieve the replica method by integrating out its couplings. With the tuned variance, the model's dominant contributions can be controlled with solvable property. In this meaning, the disorder averaging is kind of a trick \cite{syktrick}.\\

Thirdly, for both the model itself and its effective action, SYK model actually preserves the $O(N_s)$ symmetry. However, the proposed model transmutation to matrix model implied by the replica method suggests that the disorder-averaged SYK model naturally reserves a holographic nature with strong couplings in the IR regime.\\

Explicitly, SYK model arrives at $AdS_2/CFT_1$ holography through the emergent conformal symmetry fitted by $SL(2,\mathbb{R})$ group.\\

As is known, the idea of holography was inspired by black hole's area-dependent entropy instead of volume-dependent, and this idea was then generalized to the universal correspondence between gauge theory and gravity. That is, any theory of holography must be related with black hole problems. One manifest success of SYK model is its well match with black hole's gravitational mode through the time dependence of the model's OTOCs.\\

Fourthly, although the SYK model itself is an essential quantum mechanical spin liquids system, there are disagreement from the quantum property of the effective theory of it. Besides the structural similarity with vector models, we further point out the coupling correspondence with them.\\

Generally speaking, SYK model is much simpler than Super Yang-Mills (SYM) theory and tensor model, and based on the correspondence with vector model as well as the model transmutation, it is like a hybrid of these two models.\\

However, from the renormalization group theory \cite{peskin}\renewcommand{\thefootnote}{\fnsymbol{footnote}}\footnote[1]{Chapter 12}\renewcommand{\thefootnote}{\arabic{footnote}}\cite{wilson}\cite{coleman}\renewcommand{\thefootnote}{\fnsymbol{footnote}}\footnote[3]{Chapter 5}\renewcommand{\thefootnote}{\arabic{footnote}}, the property of universality tells us that the fundamental laws of physics are not determined by the specified formula of the theory. Instead, they depend only on the basic symmetry principles of the theory because the property of RG flow and its fixed points rely only on the basic symmetries of Lagrangians (Hamiltonians) continuously varied by them.\\

Phrased another way, different symmetries lead to different theories, and symmetry matters the most in a theory no matter what exact formula it takes.\\

In this level, no matter whether the effective action of SYK model is about the real theory of quantum gravity or not, the emergent conformal symmetry it preserves matters a lot.\\

Moreover, more symmetries mean more physics. Other than the vector and matrix models, Gross-Neveu (GN) model is an $(1+1)$-dimensional field theory of fermions with discrete chiral symmetry as well as an $SU(N_f)$ symmetry \cite{aneesh},
\begin{eqnarray}\label{gnms1}
\mathcal{L} \sim \sum_i^{N_f}  \bar{\psi}_i i\gamma^{\mu}\partial_{\mu}\psi_i + \lambda (\sum_i^{N_f} \bar{\psi}_i \psi_i)^2,
\end{eqnarray}

Remarkably, its effective theory can be semiclassically expressed as,
\begin{eqnarray}\label{gnms}
S_{_{\textrm{eff}}}\sim N_f \cdot f(g^2N_f),
\end{eqnarray}
with the fixed $g^2N_f$ resembles the effective coupling in matrix model.\\

From both the symmetry side and the coupling correspondence, compared with the original 2-index SY model and the matrix/vector models, GN model is much more closer to SYK model \cite{peng}. Thus, the relation between them is worthy to be investigated.

\acknowledgments
The authors have been supported by Australian Research Council Discovery Project DP190101529 and National Natural Science Foundation of China (Grant No.11775177).

\end{document}